\documentstyle[epsfig,aps,amsmath,twocolumn,floats]{revtex}

\begin{document}
\draft
\wideabs{
\title{Model of a fluid at small and large length scales and the 
hydrophobic effect}
\author{Pieter Rein ten Wolde, Sean X. Sun and David Chandler}
\address{Department of Chemistry, University of California, Berkeley,
California 94720}
\maketitle

\begin{abstract}
We present a statistical field theory to describe large length scale
effects induced by solutes in a cold and otherwise placid liquid. The
theory divides space into a cubic grid of cells. The side length of
each cell is of the order of the bulk correlation length of the bulk
liquid. Large length scale states of the cells are specified with an
Ising variable. Finer length scale effects are described with a
Gaussian field, with mean and variance
affected by both the large length scale field and by the constraints
imposed by solutes. In the absence of solutes and corresponding
constraints, integration over the Gaussian field yields an effective
lattice gas Hamiltonian for the large length scale field. In the
presence of solutes, the integration  adds
additional terms to this Hamiltonian. We identify these terms
analytically. They can provoke large
length scale effects, such as the formation of interfaces and
depletion layers. We apply our theory to 
compute the reversible work to form a bubble in liquid water, as a
function of the bubble radius. Comparison with molecular
simulation results for the same function indicates that the theory is
reasonably accurate. Importantly, simulating the large length scale
field involves binary arithmetic only. It thus provides a
computationally convenient scheme to incorporate explicit solvent
dynamics and structure in simulation studies of large molecular
assemblies.
\end{abstract}

\pacs{PACS numbers: 61.20.-p, 61.20.Gy, 68.08.-p, 82.70.Uv, 87.15.Aa}
}

\newpage

\section{Introduction}

We have constructed a tractable model for describing density fluctuations in
a cold liquid at both small and large length scales. The model allows
us to analyze at a microscopic level the effects of solvated surfaces and
large molecular assemblies, perhaps of biophysical relevance. This paper
presents the model and demonstrates its tractability.

A cold liquid is a fluid that is well below the critical
temperature. Water at ambient conditions is an example. When
unperturbed, it will have no significant large length scale
fluctuations. It is nearly incompressible.  When perturbed by a
sufficiently extended surface, however, a cold liquid may exhibit
large length scale fluctuations, akin to a phase transition, in the
vicinity of the surface. This phenomenon occurs when another phase is
close to coexistence with the liquid, and when interactions with the
surface favors the other phase over the liquid. This coincidence of
conditions is pertinent, for instance, to hydrophobic effects. In
particular, water at ambient conditions lies close to coexistence with
its vapor. Further, the demixing of oil and water and the associated
large oil-water surface tension indicates that a large hydrophobic
(i.e., oily) surface favors vapor over liquid water.

 Indeed, Lum,
Chandler and Weeks (LCW ) \cite{Lum99} have demonstrated that oily
surfaces extending over 1 nm or more will nucleate a layer of depleted
water density and concomitant large length scale correlations. In
contrast, perturbations from smaller hydrophobic surfaces, less than 1
nm across, do not nucleate such a drying layer and affect only small
length scale fluctuations in the liquid. Since hydrophobicity vividly
manifests the interplay and competition between small and large length
scale fluctuations in a cold liquid, we have chosen in this paper to
focus attention on it. One benefit of our analysis is an understanding
of the results of LCW theory from a perspective that is numerically
simpler and physically more transparent than the original LCW
development. Generalizations of our approach to other phenomena,
including the effects of strong associative interactions between
solutes and solvent, should be apparent.

The main idea of our approach is to create a statistical field theory
where the molecular density field is decomposed into two parts. One
part varies on large length scales only. The other varies on small
length scales.  For a cold fluid that is homogeneous and therefore
nearly incompressible, the large length scale field is nearly constant
and equal to the mean density of the bulk liquid. Even for this
homogeneous case, however, small length scale fluctuations are always
present. To a remarkable extent~\cite{Hummer96,Crooks97}, the
statistics of these fluctuations is Gaussian with a variance
determined by the structure factor of the bulk liquid.  Accurate
molecular theories of solvation and liquid structure at small length
scales -- the Percus-Yevick equation for hard sphere
fluids~\cite{Percus58,Wertheim63}, the mean spherical
approximation~\cite{Lebowitz66,Wertheim71}, the
Pratt-Chandler theory of hydrophobicity~\cite{Pratt77}, the reference
interaction site model (RISM)~\cite{Chandler72,Schweizer87} -- are
consequences of such statistics~\cite{Chandler93}. This Gaussian
statistics for small length scale fluctuations is an important element
of the weight functional (or Hamiltonian) we construct.  These
fluctuations are coupled, of course, to the large length scale density
field, and they are also constrained by the presence of solutes. Due
to the coupling and constraints, the variance of the small length
scale fluctuations may differ markedly from that of the homogeneous
bulk fluid.

The Hamiltonian for our model is presented in
Sec.~\ref{sec:model}. The large length scale density field supports possible
phase coexistence and interfaces. As such, we see in that section how
the coupling between small and large length scale fields may lead to
solute induced interfaces in a cold fluid. Our treatment of this
coupling is inspired by the work of Weeks and his
coworkers~\cite{Lum99,Weeks98}. They related the coupling to
unbalanced attractive forces that result from local inhomogeneities in
the fluid.  Analytical integration over the small length scale field
is possible due to its Gaussian statistics. The integration yields an
effective Hamiltonian functional for the large length scale field. In
Sec.~\ref{sec:theory}, we describe how this integration can be used to study
solvation. This step also lays the foundation for a numerical scheme where
the solvent is simulated at the level of the large length scale
density field. Such a scheme involves only binary arithmetic and is
much more efficient than an atomic level simulation. In fact, it is
sufficiently efficient to make possible the study of phenomena like
self-assembly of biological structures.

 In Sec.~\ref{sec:comparison}, we discuss the
results of our treatment in various limits. In the absence of any
solutes, the effective Hamiltonian for the large length scale density
corresponds to the lattice gas model~\cite{Chandler87}. In the
presence of solutes that are small in size and number, only those density
fluctuations at small length scales are relevant, and our model
reduces to the Gaussian model of Pratt and Chandler~\cite{Pratt77},
and the closely related information theory approach of Hummer, Pratt
and coworkers~\cite{Hummer96}.  In the presence of large solutes, a
mean field approximation to our model coincides with the LCW theory
\cite{Lum99}.

A numerical application is given in Sec.~\ref{sec:application}. We
first show how the parameters in our model can be estimated from
experimentally accessible quantities. We then explicitly treat the
solvation of an ideal hydrophobic sphere in water and compare our
results with those of an atomistic simulation~\cite{Huang01_1}. Finally, in
Sec.~\ref {sec:discussion}, we discuss implications and possible
extensions  of this work.

\section{Model}

\label{sec:model} 
\begin{figure}[t]
\epsfig{figure = 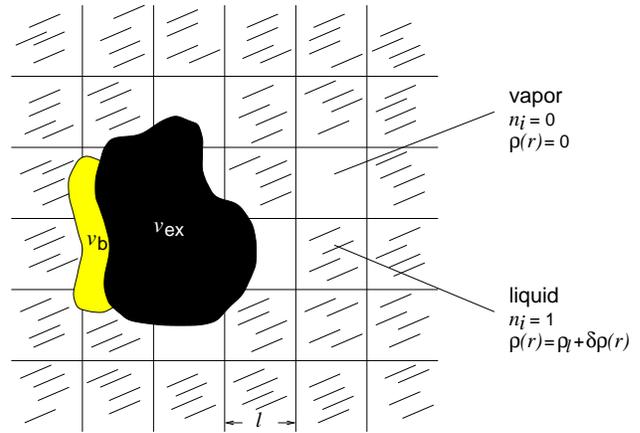,width=8.25cm}
\vspace*{0.425cm}
\caption{\label{fig:sketch}
Sketch of our model of a cold liquid in the presence of a
solute. The solute excludes a volume $v_{\mbox{\tiny ex}}$ from the
solvent (black region) and has a hydrophilic patch of strong
associative interactions with the solvent; the patch imposes a
constraint on the solvent density in the volume $v_{\mbox{\tiny b}}$
(gray area). The solvent is divided into cells of width $l$; each cell
is either filled with liquid ($n_{i}=1$) or vapor ($n_{i}=0$). The
field $n_{i}$ describes density fluctuations on length scales larger
than the lattice spacing. This field supports phase
transitions. Density fluctuations on length scales smaller than the
lattice spacing are described by the field $\delta \rho ({\bf
r})$. This field describes molecular detail such as the highly
oscillatory profiles for the average density near small solutes. We
thus write the density as $\rho ({\bf r}_i)=n_{i}\rho _{l}+\delta \rho
({\bf r}_i)$. The field $ \delta \rho ({\bf r})$ is assumed to obey
Gaussian statistics.
}
\end{figure}

Figure~\ref{fig:sketch} illustrates the essential features
of a cold fluid in the presence of a solute. The solute is of arbitrary size
and shape. If it is small, the solvent will wet its surface. In contrast, if
the solute is large with extended hydrophobic surfaces, solvent density near
the solute will be depleted relative to the density of the bulk liquid,
$\rho _{l}$. This drying-like phenomenon occurs because the solvent
experiences significant unbalanced attractive forces near the hydrophobic
surface. These forces induce depletion. The solute can also have patches of
associative interactions. Adjacent to those patches, the molecular density
of the solvent will be close to or perhaps greater than that of the bulk
liquid.

In our description of solvation, we make a distinction between strong
forces and weak solvent-solute forces. The repulsive nearly hard core
interactions between solute and solvent molecules are strong
forces. So too are associative interactions between solute and
solvent. On the other hand, dispersion interactions between solute and
solvent molecules are weak forces. In some cases, electrostatic forces
are weak forces. Weak interactions are described in our treatment by
an interaction potential acting between the solute and the solvent
density. In contrast, strong forces are treated according to the
constraints they impose upon the solvent density fluctuations. For
example, the effect of a solute repulsive core is mainly to exclude
solvent from a volume, $v_{{\rm ex}}$ in Fig.\ref {fig:sketch}. The
effect of these forces may be described as a constraint permitting only
those fluctuations in the solvent density, $\rho \left( {\bf r
}\right)$, that leave $v_{{\rm ex}}$ empty of solvent, i.e., $\rho
({\bf r} )=0$ for ${\bf r}\in v_{{\rm ex}}$~\cite{Chandler93}. Similarly,
associative interactions may cause $n$ water molecules to be bound
within a specific region close to the solute. In Fig.\ref{fig:sketch},
this region is $v_{{\rm b}}$. This effect can be treated by
constraining the integral of $\rho ({\bf r})$ over the volume $v_{{\rm
b}}$ to equal $n$~\cite{Lum99}. We do not apply this latter idea in
the current paper, although the methods by which it can be implemented
should be clear from our treatment of the former.

Since the fluid is assumed to be cold, the regions of gas or vapor can
be clearly distinguished from those of liquid. The density of the
vapor is typically orders of magnitude smaller than that of the
liquid. In such a situation, it is natural to divide space into a grid
of cells, where each cell contains either gas or liquid. We use cubic
cells, and take the distance
across each cell, $l$, to be comparable to the bulk
liquid correlation length, $ \xi $.  In that case, a binary choice
of states within a cell, either gas or liquid, provides a reasonable
coarse grained rendering of likely configurations of the fluid. We can
thus define a field, $n_{i}$, that takes on the value of $1$ if
 cell $i$ contains liquid and $0$ if it contains gas. The molecular
density we associate with this field is $ n_{i}\rho _{l}$. This
field, $n_{i}$, or equivalently $n_{i}\rho _{l}$, is the large length
scale field in our model. It can be used together with a second field,
$\delta \rho ({\bf r})$, to describe the density on length scales
both larger and smaller than $l$. In particular, for positions ${\bf
r }$ within cell $i,$ ${\bf r}_{i},$ we write the net density as
\begin{equation}
\rho ({\bf r}_{i})=n_{i}\rho _{l}+\delta \rho ({\bf r}_{i}).
\end{equation}
All of space is spanned by the set of ${\bf r}_{i}$, i.e., $\int d{\bf r}
\equiv \sum_{i}\int d{\bf r}_{i}$.

While the field $n_{i}$ is binary and can be used to describe a liquid-gas
phase transition, the field $\delta \rho ({\bf r})$ has a very different
character. It supports neither phase transitions nor interfaces, but it
does describe small length scale structures such as those manifesting the
granularity of the solvent (e.g., the oscillatory profiles of the average
liquid density in the vicinity of a small solute). It must be possible,
therefore, that $\delta \rho ({\bf r})$ can take on a variety of values. As
indicated in the Introduction, it is a reasonable approximation to adopt the
simplest possible statistics for this field. Namely, we assume it is
Gaussian and define its variance to be
\begin{equation}
\chi [{\bf r}_{i},{\bf r}_{j}^{\prime };\{n_{k}\}]=\langle \delta \rho ({\bf 
r}_{i})\delta \rho ({\bf r}_{j}^{\prime })\rangle _{\{n_{k}\}}.
\label{eq:chiG}
\end{equation}
Here, $\left\langle ...\right\rangle _{\{n_{k}\}}$ indicates the
ensemble average over density fluctuations for a given configuration
of the field $ n_{k}$. The dependence upon $n_{k}$ is
significant. When $n_{k}=1$ for all $ k,$ corresponding to a cold
liquid with absolutely no large length scale fluctuations, $\delta
\rho ({\bf r}_{i})$ has zero mean, and its variance reduces to the
response function of the bulk fluid,
\begin{equation}
\chi ({\bf r}_{i},{\bf r}_{j}^{\prime };\rho _{l})=\rho _{l}\delta ({\bf r}
_{i}-{\bf r}_{j}^{\prime })+\rho _{l}^{2}h(|{\bf r}_{i}-{\bf r}_{j}^{\prime
}|;\rho _{l}),
\end{equation}
where $h(|{\bf r}_{i}-{\bf r}_{j}^{\prime }|;\rho _{l})+1$ is the
radial distribution function of the uniform fluid at density $\rho
_{l}$. On the other hand, within a cell that contains vapor ($n_i=0$),
small length scale fluctuations are very small. Our model
employs the approximation that $\delta \rho ({\bf r} _{i})=0$ whenever
$n_{i}=0$. Therefore, we imagine that $\delta \rho ({\bf r})$ is a
Gaussian field, with a weight functional being that of the bulk fluid,
but constrained to be zero whenever $n_i=0$. The response function for
such a field is~\cite{Chandler93}
\begin{eqnarray}
\label{eq:chi_full}
\chi [{\bf r}_{i},{\bf r}_{j}^{\prime };\{n_{k}\}]&=&\chi ({\bf r}_{i},{\bf r}
_{j}^{\prime };\rho _{l}) \nonumber\\[0.25cm]
&&-\sum_{k}\sum_{l}\int d{\bf r}^{\prime \prime}_{k}\int d{\bf r}
_{l}^{\prime \prime \prime }\chi ({\bf r}_{i},{\bf r}^{\prime
\prime}_{k};\rho _{l}) \nonumber \\[0.25cm]
&&\hspace*{0.35cm}
\times \: \chi_{\rm g}^{-1}[
{\bf r}^{\prime \prime}_{k},{\bf r}_{l}^{\prime \prime
\prime};\{n_{k}\}] \: \chi ({\bf r}_{l}^{\prime \prime
\prime },{\bf r}_{j}^{\prime };\rho _{l}).
\end{eqnarray}
Here, 
\begin{eqnarray}
\chi _{\rm g}[{\bf r}_{i},{\bf r}_{j}^{\prime };\{n_{k}\}] &=&\chi ({\bf r}_{i},
{\bf r}_{j}^{\prime };\rho _{l})\mbox{     if $n_i = n_j = 0$,}  \nonumber \\
&=&0\mbox{     otherwise,}
\end{eqnarray}
is the $({\bf r}_i,{\bf r}^{\prime}_j)$ element of the matrix ${\boldsymbol
 \chi}_{\rm g}$.
Similarly, the matrix with elements $\chi _{\rm g}^{-1}[{\bf r}_{i},{\bf
r}_{j}^{\prime };\{n_{k}\}]$ is also 
non-zero only when $n_{i}=n_{j}=0$. In that space, where $n_{i}=0$,
${\boldsymbol \chi}_{\rm g}^{-1}$ is the
inverse of ${\boldsymbol \chi}_{\rm g}$. These
relations project the matrix with elements $\chi^{-1}[{\bf r}_{i},{\bf r}_{j}^{\prime };\{n_{k}\}]$
onto the space of cells for which $n_{i}=1$. We adopt these relations to
define our model of Gaussian statistics for $\delta \rho ({\bf r})$.

With the lattice spacing as large as the bulk correlation length, the
field $n_i \rho_l$ is nearly incompressible. This means that, in the
absence of any strong perturbations on the fluid, the field $n_{i}$ is
essentially constant. In this case of the unperturbed (i.e., uniform)
fluid, density fluctuations are described almost entirely by the field
$\delta \rho ({\bf r})$. The compressibility of the uniform fluid is
contained in the variance for $ \delta \rho ({\bf r})$. In this
context, consider the behavior of the bulk liquid structure factor,
$\int d({\bf r}-{\bf r}^{\prime })\chi ({\bf r}, {\bf r}^{\prime
};\rho _{l})\exp [i{\bf k}\cdot ({\bf r}-{\bf r}^{\prime })]. $ The
long-wave length limit of the structure factor is proportional to the
bulk compressibility. It approaches this limit with a plateau. In
particular, for $k$ values smaller than some finite wave-vector
$k_{{\rm c}}$ , the structure factor is essentially constant. The grid
spacing we use to define large length scales coincides with $l\sim
2\pi /k_{{\rm c}}$.

To within a physically irrelevant metric factor, the partition function for
our model is
\begin{eqnarray}
\label{eq:Xi}
\Xi &=&\sum_{\{n_{i}\}}\int {\cal D} \delta \rho ({\bf r}) \: C[\left\{
n_{k}\right\},\delta \rho ({\bf r})] \: \nonumber \\
&&\times \: \exp \left(-\beta H[\left\{
n_{k}\right\} ,\delta \rho ({\bf r})]\right),
\end{eqnarray}
where $\int {\cal D}\delta \rho ({\bf r})=\int \Pi _{i}{\cal D}\delta \rho (
{\bf r}_{i})$ denotes the functional integration over the small length scale
field, $H[\left\{ n_{k}\right\} ,\delta \rho ({\bf r})]$ is the Hamiltonian
as a functional of both $n_{i}$ and $\delta \rho ({\bf r}),$ and $\beta
^{-1} $ is Boltzmann's constant time temperature, $k_{{\rm B}}T$. The
quantity $C[\left\{ n_{k}\right\} ,\delta \rho ({\bf r})]$ is a constraint
functional. It has unit weight when the field $\delta \rho ({\bf r})$ together with
$\left\{ n_{i}\right\} $ satisfy whatever constraints are imposed by strong
forces, and it is zero otherwise. Since $\{n_i\}$ and $\delta
\rho({\bf r})$ have greatly different character, the summation and
integration in Eq.~(\ref{eq:Xi}) do not redundantly count configuration
space to any significant degree.

In our model, there are three principal contributions to the Hamiltonian, $
H[\left\{ n_{k}\right\} ,\delta \rho ({\bf r})]$. One is a lattice gas
Hamiltonian for the large length scale field,
\begin{equation}
\label{eq:H_L}
H_{{\rm L}}[\left\{ n_{k}\right\} ]=-\mu \sum_{i}n_{i}-\epsilon
\sum_{<i,j>}n_{i}n_{j}.
\end{equation}
Here, $\mu $ is the imposed chemical potential, the sum labeled with $
\left\langle ij\right\rangle $ is over all nearest neighbor pairs of cells,
and the interaction parameter $\epsilon $ determines the energetic cost of
creating a vapor-liquid interface. Importantly, the lattice gas model
supports phase transitions and sustains gas-liquid interfaces.

A second contribution to the Hamiltonian
ensures the Gaussian weight for the small length scale field. From the
principle of equipartition, this contribution must be
\begin{equation}
\frac{k_{{\rm B}}T}{2}\sum_{i,j}\int d{\bf r}_{i}\int d{\bf r}_{j}^{\prime
} \: \delta \rho ({\bf r}_{i}) \:\chi ^{-1}[{\bf r}_{i},{\bf r}_{j}^{\prime
};\{n_{k}\}] \: \delta \rho ({\bf r}_{j}^{\prime }).
\end{equation}

A third contribution to the Hamiltonian gives the coupling between the $
n_{i} $ and $\delta \rho ({\bf r})$ fields arising from unbalanced forces.
According to the arguments provided by Weeks and coworkers for simple fluids
\cite{Lum99,Weeks98}, the unbalancing potential acting on $n_{i}$ for a
simple fluid is well estimated by $-2a\langle \overline{\delta \rho ({\bf r})
}\rangle $. Here, $a\rho _{l}^{2}$ is the energy density of the uniform
liquid at density $\rho _{l}$, and the overbar denotes a coarse-graining of
the density fluctuation $\delta \rho ({\bf r})$ over a length scale
comparable to the bulk correlation length. Based upon this estimate, we
write the contribution to the Hamiltonian from unbalanced forces as
\begin{equation}
-2 \epsilon ^{\prime }\sum_{<i,j>}\int d{\bf r}_{i} \: \delta \rho ({\bf r}_{i})
\: \frac{n_{j}-1}{\rho _{l}l^{3}}.
\end{equation}
For simple fluids with only one energy and length scale, $\epsilon
^{\prime }=\epsilon $.  This equality implies that the lattice gas
parameters $\epsilon$ and $l$ are sufficient to determine both the
surface tension and the energy density of the liquid. For more complex fluids, including water, the
effects of orientational degrees of freedom may introduce multiple
microscopic length scales, and as a result,  $\epsilon
^{\prime }$ could differ from $\epsilon $. This possibility was ignored in
Ref.\cite{Lum99}, but will be examined in Sec.~\ref{sec:discussion}.

By combining all three contributions, we arrive at our final result for the
Hamiltonian of our model. It is
\begin{eqnarray}
 H[\left\{ n_{k}\right\},\delta \rho ({\bf r})] & = &
 H_{{\rm L}}[\left\{
n_{k}\right\} ]-2 \epsilon^{\prime }\negthickspace \sum_{<i,j>}\int d{\bf r}_{i} \,
\delta \rho({\bf r}_{i}) \, \frac{n_{j}-1}{\rho _{l}l^{3}} \nonumber \\[0.25cm]
&&+ \: \frac{k_{\rm B}T}{2}\sum_{i,j}\int d{\bf r}_{i}\int d{\bf
r}_{j}^{\prime} \:\nonumber \\[0.15cm]
&&\hspace*{1.0cm} 
\times \: \delta \rho ({\bf r}_{i}) \: \chi ^{-1}[{\bf
r}_{i},{\bf r}_{j}^{\prime};\{n_{k}\}] \: \delta \rho ({\bf
r}_{j}^{\prime }) \nonumber \\[0.25cm]
&&+ \: H_{\mbox{\tiny norm}}[\left\{ n_{k}\right\}],  \label{eq:H}
\end{eqnarray}
where
\begin{eqnarray}
H_{{\mbox{\tiny norm}}}[\left\{ n_{k}\right\} ] &=&
\frac{k_{{\rm B}}T}{2}
\sum_{\left\langle i,k\right\rangle }\sum_{\left\langle j,l\right\rangle
}\int d{\bf r}_{i}\int d{\bf r}_{j}^{\prime } \:\nonumber\\[0.15cm]
&&\hspace*{2.0cm} 
\times \: \phi _{k}
\: \chi
[{\bf r}_{i},{\bf r}_{j}^{\prime };\{n_k\}] \: \phi _{l} \nonumber \\[0.25cm]
&&+\: k_{\rm B}T\ln \sqrt{\mbox{det}{\boldsymbol \chi}},
\end{eqnarray}
and
\begin{equation}
\phi _{j}=2 \beta \epsilon^{\prime } \: \frac{n_{j}-1}{\rho _{l}l^{3}}.
\end{equation}
Here, the quantity $\mbox{det} {\boldsymbol \chi}$ is the determinant of the matrix
with elements $\chi[{\bf r}_i,{\bf r}^{\prime}_j;\{n_k\}]$.  The last term in
Eq.~(\ref{eq:H}), $H_{{\mbox{\tiny norm}}}[\left\{ n_{k}\right\} ]$,
provides a normalization constant for the functional integration over
$ \delta \rho ({\bf r})$. When there are no strong forces, so that
the constraint functional $C[\left\{ n_{k}\right\} ,\delta \rho ({\bf
r})]$ is simply unity, the effective Hamiltonian for the $n_{i}$
field should be exactly the lattice gas Hamiltonian,
Eq.~(\ref{eq:H_L}). The last term in Eq.~(\ref{eq:H}) ensures that the
integration over $\delta \rho ({\bf r})$ for this case will indeed
yield this result.

\section{Theory of solvation}

\label{sec:theory} The excess chemical potential of a solute, $\Delta \mu ,$
is given by~\cite{Rowlinson82}
\begin{equation}
\beta \Delta \mu =-\ln \frac{\Xi _{{\rm S}}}{\Xi }=-\ln \langle \exp (-\beta
U_{{\rm S}})\rangle _{0}.
\end{equation}
Here $\Xi $ is the partition function for the unperturbed solvent and $\Xi _{
{\rm S}}$ is the partition function for the system in the presence of a
(fixed) solute. The energy $U_{{\rm S}}$ is the energy of interaction
between the solute and the solvent molecules and the subscript $0$ denotes
an ensemble average over the unperturbed solvent.

For simplicity, we will consider the solvation of an ideal hydrophobic
solute in water -- a particle that excludes water from a region
$v_{{\rm ex} }$, but has no other interactions with the solvent. A
hard sphere is an example of an ideal hydrophobic solute. It excludes
solvent from a volume $v_{{\mbox{\tiny ex}}}=(4/3)\pi R^{3}$, where
the radius $R$ is the distance of closest approach between water and
solute. The partition function of the system in the presence of such a
solute is equal to the partition function of the unperturbed solvent,
but with the constraint that no solvent exists inside the excluded
volume. In other words, the constraint functional for this case is
\begin{equation}
\label{eq:Cfull}
C[\left\{n_{k}\right\},\delta \rho ({\bf r})]=\prod_{{\bf r}_{i}\in
v_{\mbox{\tiny ex}}}\delta [n_{i}\rho _{l}+\delta \rho ({\bf r}_{i})].
\end{equation}
Accordingly, the partition function in the presence of an ideal hydrophobic
solute is
\begin{eqnarray}
\Xi _{{\rm S}}&=&\sum_{\{n_{i}\}}\int \prod_{i}{\cal D}\delta \rho ({\bf r}
_{i}) \left\{ \prod_{{\bf r}_{i}\in v_{\mbox{\tiny ex}}}\delta
[n_{i}\rho _{l}+\delta \rho ({\bf r}_{i})]\right\} \nonumber\\[0.25cm]
&& \times \: \exp \left( -\beta
H[\left\{ n_{k}\right\},\delta \rho ({\bf r})]\right). \label{eq:X_s} 
\end{eqnarray}

For an ideal hydrophobic solute, the ratio of partition functions $\Xi _{
{\rm S}}/\Xi $, equals the probability of
observing no solvent molecules inside the volume $v_{\mbox{\tiny ex}}$.
Equivalently, it corresponds to the
probability of observing a cavity of volume $v_{\mbox{\tiny ex}}$
inside the solvent; it is also equal to the probability that a solute
can be inserted into the solvent without creating any overlap with the
solvent molecules.  The excess chemical potential of an ideal solute
could be obtained by imposing an alternative constraint,
\begin{equation}
\label{eq:Cob}
C[\left\{ n_{k}\right\},\delta \rho ({\bf r})]=\delta \left[
\int_{{\bf r} \in v_{\mbox{\tiny ex}}} d{\bf r} \rho({\bf r}_i)\right].
\end{equation}
Were our treatment completely consistent with the particulate nature
of matter, the two constraints, as given by Eq.~(\ref{eq:Cfull}) and
Eq.~(\ref{eq:Cob}), would be equivalent. But in fact, Gaussian
statistics for the field $\delta \rho({\bf r})$ cannot be completely
consistent with this nature of matter, and the two constraint
functionals will yield somewhat different results.

To evaluate the partition function, Eq.~(\ref{eq:X_s}), it is convenient to 
rewrite the constraint functional with the Fourier representation of delta
functions. Namely,
\begin{eqnarray}
\Xi _{{\rm S}} &=&\sum_{\{n_{i}\}}\int \prod_{i}{\cal D}\delta \rho ({\bf r}
_{i})\int \prod_{i}{\cal D}\psi ({\bf r}_{i})  \nonumber \\
&&\times \: \exp \Bigg(-\beta H[\left\{ n_{k}\right\},\delta \rho ({\bf r}
)] \nonumber \\
&& + \: i\sum_{i}\int_{{\bf r}_{i}\in v_{\mbox{\tiny ex}}}d{\bf r}_i  \psi ({\bf 
r}_{i}) \: [n_{i}\rho _{l} + \delta \rho({\bf r}_i)]\Bigg) .
\end{eqnarray}
Functional integration over both $\delta \rho ({\bf r})$ and $\psi
({\bf r})$ is now straightforward, yielding
\begin{equation}
\Xi_{{\rm S}}=\sum_{\{n_{i}\}}\exp \left( -\beta H[\left\{ n_{k}\right\}
]\right) ,
\end{equation}
where the effective Hamiltonian $H[\left\{ n_{k}\right\} ]$ is
\begin{eqnarray}
H[\left\{ n_{k}\right\} ] &=&H_{{\rm L}}[\left\{ n_{k}\right\} ]
\nonumber \\[0.25cm]
&& + \: \frac{k_{
{\rm B}}T}{2}\sum_{i,j}\int_{{\bf r}_{i}\in v_{\mbox{\tiny ex}}} \negthickspace d{\bf r}
_{i}\int_{{\bf r}_{j}^{\prime }\in v_{\mbox{\tiny ex}}} \negthickspace d{\bf r}
_{j}^{\prime } \: (n_{i}\rho _{l}+f({\bf r}_{i}))  \nonumber \\[0.25cm]
&&\hspace*{1.7cm} 
\times \: \chi _{\mbox{\tiny in}}^{-1}[{\bf r}_{i},{\bf r}_{j}^{\prime
};\{n_{k}\}] \: (n_{j}\rho _{l}+f({\bf r^{\prime }}_{j})) \nonumber \\[0.25cm]
&&+ \: k_{{\rm B}}T\ln \sqrt{
\mbox{det}{\boldsymbol \chi}_{\mbox{\tiny in}}},
\end{eqnarray}
with
\begin{equation}
\label{eq:f_r}
f({\bf r}_{i})\equiv \beta \epsilon^{\prime} \sum_{j}\int d{\bf r}
_{j}^{\prime }\sum_{k({\rm nn}j)}\frac{n_{k}-1}{\rho _{l}l^{3}}\: \chi [{\bf r}
_{i},{\bf r}_{j}^{\prime };\{n_{k}\}].
\end{equation}
Here, ${\boldsymbol \chi}_{\mbox{\tiny in}}$ has elements $\chi [{\bf r}_{i},{\bf r}
_{j}^{\prime };\{n_{k}\}]$ for ${\bf r}_{i}$ and ${\bf r}_{j}^{\prime }$
both within the excluded volume, and a sum over $k({\rm nn}j)$ is over cells
$k$ that are nearest neighbors to cell $j$.

The evaluation of $H[\left\{ n_{k}\right\} ]$ requires the calculation
of various integrals and matrix inverses. These quantities can be
conveniently estimated to a good approximation by exploiting the fact
that the lattice spacing is on the order of the bulk correlation
length. In particular, since the bulk correlation function, $\chi
({\bf r}_{i},{\bf r}_{j}^{\prime };\rho _{l})$ vanishes quickly for
$|{\bf r}-{\bf r}^{\prime}|$ larger than that length, $\chi [{\bf r}_{i},{\bf
r}_{j}^{\prime };\{n_{k}\}]$ as given by Eq.~(\ref{eq:chi_full}), can be
approximated by
\begin{eqnarray}
\chi [{\bf r}_{i},{\bf r}_{j}^{\prime };\{n_{k}\}] &=&\chi ({\bf r}_{i},{\bf 
r}_{j}^{\prime };\rho _{l}),\quad {\rm for}\quad n_{i}=n_{j}=1 \nonumber\\
&=&0,\quad {\rm otherwise}.
\end{eqnarray}
Furthermore, the relatively large size of the cells allows us to
restrict the sum in Eq.~(\ref{eq:f_r}) to the $i=j$ term, and to take
the integral over all space, rather than over one cell. As such, we
arrive at a much simplified form for $f({\bf r}_{i})$, and therefore,
\begin{equation}
f_{i}\equiv \int_{{\bf r}_{i}\in v_{\mbox{\tiny ex}}} \negthickspace d{\bf r}_{i}f(
{\bf r}_{i})= n_{i}\:v_{i}\:\epsilon^{\prime }\:\kappa \:\frac{\rho _{l}}{l^{3}}
\sum_{k({\rm nn}i)}(n_{k}-1).
\end{equation}
Here $\kappa $ is the isothermal compressibility of the uniform fluid, which
is related to the response function via $\kappa =\beta \widehat{\chi }
(0)/\rho _{l}^{2}$, where $\widehat{\chi }(0)$ is the long-wave length limit
of the Fourier transform of the structure factor. Note that $f_{i}$ is zero,
when cell $i$ is not liquid, i.e., when $n_{i}=0$.

Finally, with $\chi ({\bf r}_{i},{\bf r}_{j}^{\prime };\rho _{l})$ provided
as input into the theory, we must choose a set of basis functions that span
the space of the excluded manifold. This allows us to perform the inversion
of $\chi _{\mbox{\tiny in}}[{\bf r}_{i},{\bf r}_{j}^{\prime };\{n_{k}\}]$ in
the representation prescribed by that basis. We use the approximation of one
basis function spanning the excluded volume and to take $\chi_{\mbox{\tiny
 in}}[{\bf r}_i,{\bf r}_j';\{n_k\}]=\chi({\bf
 r}_i,{\bf r}_j';\rho_l)$ for all cells $i$ inside the excluded
volume~\cite{note1TenWolde01}. We then
arrive at our principal result:
\begin{eqnarray}
H[\left\{ n_{k}\right\} ] &=&H_{{\rm L}}[\left\{ n_{k}\right\} ]
\nonumber \\
&&+ \: k_{{\rm B}
}T\sum_{i,j({\rm occ})}\frac{n_{i}\left[ \rho _{l}v_{i}+f_{i}\right] \left[
\rho _{l}v_{j}+f_{j}\right] n_{j}}{2\sigma _{v_{\mbox{\tiny
ex}}}}\nonumber\\
&&+ \: k_{{\rm B}
}T\ln \sqrt{2\pi \sigma _{v_{\mbox{\tiny ex}}}};  \nonumber \\[0.35cm]
&\equiv &H_{{\rm L}}[\left\{ n_{k}\right\} ]+H_{{\rm S}}[v_{\mbox{\tiny ex}
};\{n_{k}\}].  \label{eq:H_fs}
\end{eqnarray}
Here the sum over $i,j({\rm occ})$ is over over cells $i$ and $j$ that are
occupied by the solute; $v_{i}$ is the volume occupied by the solute in cell
$i,$ and
\begin{equation}
\sigma _{v_{\mbox{\tiny ex}}}=\int_{v_{\mbox{\tiny ex}}}d{\bf r}\int_{v_{
\mbox{\tiny ex}}}d{\bf r}^{\prime }\chi ({\bf r},{\bf r}^{\prime };\rho
_{l}).
\end{equation}
In the one-basis set approximation for $\chi[{\bf
r}_i,{\bf r}_j^{\prime};\{n_k\}]$, employed to arrive at
Eq.~(\ref{eq:H_fs}), the effect of the constraint
functional as given by Eq.~(\ref{eq:Cfull}), reduces to that of
the constraint functional as given by Eq.~(\ref{eq:Cob}).

The term $H_{{\rm S}}[v_{\mbox{\tiny ex}};\{n_{k}\}]$ contains all the
effects of the interaction between the solute and the ideal
hydrophobic solvent. It increases with increasing solute size, if
$n_i=1$ for the cells $i$ that are occupied by the solute. The
interaction term solely arises from the constraint that is imposed on
the allowed density fluctuations of the solvent. This idea, that
solvation of a hydrophobic species is equivalent to the effect of
imposing a constraint on the solvent density, is an important feature
of our model. Interestingly, the excess chemical potential of the
solute can be obtained by averaging this interaction free energy as
follows:
\begin{eqnarray}
\beta \Delta \mu (v_{\mbox{\tiny ex}}) &=&-\ln \frac{\Xi _{\rm S}}{\Xi }
\nonumber\\[0.15cm]
&=&-\ln \frac{\sum_{\{n_{i}\}}\exp (-\beta H[\left\{ n_{k}\right\} ])}{
\sum_{\{n_{i}\}}\exp (-\beta H_{{\rm L}}[\left\{ n_{k}\right\} ])}  \nonumber
\\[0.15cm]
&=&-\ln \left\langle \exp (-\beta H_{{\rm S}}[v_{\mbox{\tiny ex}
};\{n_{k}\}]\right\rangle _{{\rm L}},  \label{eq:bdmu}
\end{eqnarray}
where $\left\langle ...\right\rangle _{{\rm L}}$ indicates the ensemble
average with the Hamiltonian $H_{{\rm L}}[\left\{ n_{k}\right\} ]=H[\left\{
n_{k}\right\} ]-H_{{\rm S}}[v_{\mbox{\tiny ex}};\left\{ n_{k}\right\} ]$.

The simple formula for $H[\left\{ n_{k}\right\} ]$, Eq.~(\ref{eq:H_fs}),
and similar formulas for more general cases, can be of enormous
practical benefit for studying self-assembly. Such studies usually
require large system sizes. In those cases, the treatment of solvent
is a primary computational bottleneck.  This is because large solutes
are solvated by huge number of solvent molecules, and an atomistic
treatment involves a correspondingly large number of coordinates and
momenta. The formula for $H[\left\{ n_{k}\right\} ],$ however, lays
the foundation for a scheme in which only the solutes are treated
explicitly at the atomic level; the solutes can be moved by a
continuous Monte Carlo or molecular dynamics scheme. The solvent, on
the other hand,
is simulated in terms of the large length scale density field, $n_{i}$. That
field can be propagated by a dynamic Monte Carlo procedure,
manipulating only binary numbers. More details of this scheme will be
discussed in a forthcoming publication~\cite{TenWolde01}.

\section{Limiting Results and Comparisons with other theories}
\label{sec:comparison}

Consider first the case where $n_{i}=1$ for all $i$. This case is physically
pertinent for solutes small in size and in number because the
concomitantly small value of $v_{{\rm ex}}$ leads to relatively small free
energetic costs for having $n_{i}=1$ for all cells $i$, even for
the cells that are occupied by the solute.
Specifically, when the solutes occupy relatively small volumes, the amount
that $H_{{\rm S}}[v_{\mbox{\tiny ex}};\left\{ n_{k}\right\} ]$ will decrease
by changing $n_{i}$ from $1$ to $0$ will not compensate the corresponding
increase in $H_{{\rm L}}[\left\{ n_{k}\right\} ]$. With $n_{i}=1$ for
all $i$, $H_{{\rm L}}[n]$ and $H_{\mbox{\tiny norm}}[\left\{ n_{i}\right\} ]$
become constants and thus irrelevant. The response function $\chi [{\bf r}
_{i},{\bf r}_{j}^{\prime };\{n_{k}\}]$ reduces to the response function of
the uniform fluid $\chi ({\bf r},{\bf r}^{\prime };\rho _{l})$. Further, the
coupling term in Eq.~(\ref{eq:H}) becomes identically zero. As such, the
Hamiltonian for the model reduces to that of the Gaussian model of Pratt and
Chandler~\cite{Pratt77,Chandler93}, namely $H\left[ \left\{ n_{i}\right\} ,\delta \rho (
{\bf r})\right] \rightarrow H_{{\rm G}}\left[ \delta \rho ({\bf r})\right] ,$
where
\begin{equation}
H_{{\rm G}}\left[ \delta \rho ({\bf r})\right] =\frac{k_{{\rm B}}T}{2}\int d
{\bf r}\int d{\bf r^{\prime }}\delta \rho ({\bf r})\chi ^{-1}({\bf r},{\bf 
r^{\prime }};\rho _{l})\delta \rho ({\bf r^{\prime }}),  \label{eq:H_G}
\end{equation}
with $\delta \rho ({\bf r})=\rho ({\bf r})-\rho _{l}$, and the response
function $\chi ^{-1}({\bf r},{\bf r^{\prime }};\rho _{l})$ being the response
function of the uniform fluid. Similarly, applying Eq.~(\ref{eq:bdmu}), we
obtain the excess chemical potential for the ideal hydrophobic solute: $
\beta \Delta \mu (v_{\mbox{\tiny ex}})=-\ln \langle \exp (-\beta H_{{\rm S}
}[v_{\mbox{\tiny ex}};\{1\}]\rangle _{{\rm L}}=\beta H_{{\rm S}}[v_{
\mbox{\tiny ex}};\left\{1\right\} ]$. But, if $n_{i}=1$ for all $i$,
then $f_{i}=0$ for all $i$, and so
\begin{equation}
\beta \Delta \mu (v_{\mbox{\tiny ex}})\simeq \rho _{l}^{2}v_{\mbox{\tiny ex}
}^{2}/2\sigma _{v_{\mbox{\tiny ex}}}+\ln \sqrt{2\pi \sigma _{v_{
\mbox{\tiny
ex}}}},  \label{eq:bdmu_G}
\end{equation}
For excluded volumes not unphysically small, 
this formula for the excess chemical potential is the solvation energy
result of Hummer and Pratt and their coworkers\cite{Hummer96,Hummer98}. We
see that it is the result of Gaussian statistics in the one basis set
approximation.

In contrast to setting $n_{i}=1$ for all $i,$ the treatment of LCW~\cite
{Lum99} assumes that $n_{i}$ can be replaced by $\left\langle
n_{i}\right\rangle ,$ where $\left\langle n_{i}\right\rangle $ is an
estimate of the mean value of $n_{i}$. In that case, Eq.~(\ref
{eq:bdmu}) gives
\begin{eqnarray}
\beta \Delta \mu (v_{\mbox{\tiny ex}}) &\simeq &H_{{\rm L}}[\left\{
\left\langle n_{k}\right\rangle \right\} ]-H_{{\rm L}}[\left\{ 1\right\}
]+\beta H_{{\rm S}}[v_{\mbox{\tiny ex}};\{\left\langle n_{k}\right\rangle
\}]\medskip  \nonumber \\[0.15cm]
&\simeq &H_{{\rm L}}[\left\{ \left\langle n_{k}\right\rangle \right\} ]-H_{
{\rm L}}[\left\{ 1\right\} ]  \nonumber \\[0.15cm]
&&+ \: k_{{\rm B}}T\sum_{i,j({\rm occ})}\left\langle n_{i}\right\rangle
\left\langle n_{j}\right\rangle \rho _{l}^{2}v_{i}v_{j}/2\sigma _{v_{
\mbox{\tiny ex}}}\nonumber \\
&&+ \: k_{{\rm B}}T\ln \sqrt{2\pi \sigma _{v_{\mbox{\tiny ex}}}},\label{eq:H_LCW}
\end{eqnarray}
where the second approximate equality follows from the first after
neglecting $f_{i}$ and $f_{j}$ in comparison with $v_{i}\rho _{l}$ and
$ v_{j}\rho _{l},$ respectively. Except in the crossover regime, this
is usually a reasonable approximation, because $\kappa \rho _{l}/\beta \sim
10^{-2}$. Within notational differences, Eq.~(\ref{eq:H_LCW}) is the
solvation energy result given by LCW theory, Eq. (9) of Ref.~\cite{Lum99}.

 The LCW formula for the mean large length scale field,
$\left\langle n_{i}\right\rangle$, can also be understood from our
model. In particular, in Eq.~(\ref{eq:H}) , let us replace in the second
term the field $\delta \rho({\bf r}_i)$ with its mean, $\left\langle
\delta \rho ({\bf r} _{i})\right\rangle$. With this replacement, the
first two terms in Eq.~(\ref {eq:H}) give the mean molecular field
$\phi_j$ acting on $n_{j}$:
\begin{eqnarray}
\phi_j &=&-\mu -\sum_{i({\rm nn}j)}\left[ \epsilon \left\langle n_{i}\right\rangle
+\epsilon ^{\prime }\int d{\bf r}_{i}\left\langle \delta \rho ({\bf r})\right\rangle
/\rho _{l}l^{3}\right]  \nonumber \\[0.15cm]
&\simeq &-\mu -\epsilon \sum_{i({\rm nn}j)}\left[\langle n_{i}\rangle
+ \overline{ \langle \delta \rho ({\bf r}_i)\rangle }/\rho_l\right]
\label{eq:nLCW}.
\end{eqnarray}
where the approximate equality follows principally from approximating
$ \epsilon ^{\prime }$ with $\epsilon$. For the coarse graining
indicated by the over-bar, we use $l$ as the coarse graining
length. Other contributions to the mean molecular field come from the
quadratic term in $\delta \rho({\bf r})$ and from $H_{\mbox{\tiny
norm}}$. These, however, are small outside the crossover regime,
either because they appear in the logarithm or because they arise from
unlikely configurations, where one neighbouring cell is filled while
another is empty. With the molecular field in Eq.~(\ref{eq:nLCW}), the
LCW self-consistent equation for $\left\langle n_{j}\right\rangle $ is
obtained. Specifically, since both $\left\langle n_{i}\right\rangle $
and $\overline{ \langle \delta \rho ({\bf r}_i)\rangle }/\rho_l$ vary
slowly in space, they may be expanded for $i$ close to $j$ about
$\left\langle n_{j}\right\rangle $ and $\overline{ \langle \delta \rho
({\bf r}_j)\rangle }/\rho_l$, respectively. Truncating the expansion of
$\left\langle n_{i}\right\rangle $ at the square gradient order, and
the expansion for $\overline{ \langle \delta \rho
({\bf r}_i)\rangle }/\rho_l $ at lowest order,
Eq.~(\ref{eq:nLCW}) gives Eq. (5) of Ref.\cite{Lum99}. Thus, the
principal results of LCW can be understood as a mean field
approximations to the model we have presented herein.

\section{Application to Hydrophobic Solvation: Free energy to hydrate
a spherical bubble}
\label{sec:application}

\subsection{Parameters}

To apply our model, experimentally accessible quantities such as the
structure factor and the surface tension of the solvent must be known. For
the structure factor, we use the data of Narten and Levy for water~\cite
{Narten71}. For the energy parameter $\epsilon $, we consider its connection
to the experimental vapor-liquid surface tension of water, $\gamma .$
Namely, since the liquid is cold, we use the low temperature relation,
\begin{equation}
\gamma =\frac{\epsilon }{2l^{2}}.
\end{equation}
With $\gamma =70\mbox{ mN/m}$, this yields $\epsilon =6.02\mbox{ }k_{{\rm B}
}T$.

For the energy parameter $\epsilon ^{\prime }$, we consider its approximate
connection with the energy density of the fluid, $-a\rho _{l}^{2}$. In
particular, 
\begin{equation}
\label{eq:a_epsp}
a=\frac{z\epsilon ^{\prime }}{2\rho _{l}^{2}\l ^{3}},
\end{equation}
where $z=6$ is the coordination number of the lattice of the cubic
lattice gas model~\cite{note2TenWolde01}. Since the lattice spacing $l$
should be on the order of the bulk correlation length, $\xi$, we have
chosen $\xi \approx \l =4.2$ \AA . We have verified that the results, such
as those given below, do not depend strongly on the precise value of
the lattice spacing by varying it between \mbox{$l=3.5$ \AA } and \mbox{$l=5.0$ \AA}. The energy density parameter $a$ was derived from the internal energy $U$, which was
obtained in the following way:
\begin{equation}
U(298\mbox{ K})\approx \Delta H_{\mbox{\tiny vap}
}(373 \mbox{ K})+\int_{373}^{298}c_{p}dT,
\end{equation}
where $\Delta H_{\mbox{\tiny vap}}$ is the heat of
vaporization and $c_{p}$ is the heat capacity. With $\Delta H_{
\mbox{\tiny
vap}}=-40.7\mbox{ kJ/mol}$, $c_{p}=75.3\mbox{ J/K/mol}$, this yields $a=566 
\mbox{ }k_{{\rm B}}T$ \AA $^{3}$ and thus $\epsilon ^{\prime }=15.2\mbox{ }
k_{{\rm B}}T$. Note that $\epsilon^{\prime}>\epsilon$. The inequality indicates
that the unbalancing potential is strong enough to induce a vapor
layer near the surface of a large hydrophobic object. In fact, the
results of the numerical application given below, suggest that the
unbalancing potential is even stronger.

For the imposed chemical potential, $\mu ,$ we use
\begin{equation}
\mu = \mu _{\mbox{\tiny coex}} + \Delta \mu \approx \mu _{\mbox{\tiny
coex}}+\Delta Pl^{3},
\end{equation}
where $\mu _{\mbox{\tiny coex}}=-3\epsilon $ is the chemical potential at
coexistence, and $\Delta P$ is the difference between the pressure at
ambient conditions and the pressure at coexistence. With $\Delta P=1.0\times
10^{5}\mbox{ Pa}$, the the chemical potential change is $\Delta \mu =
5.50 \times 10^{-4}\mbox{ }k_{{\rm B}}T$. Note that $\Delta \mu$ is
very small. Water at ambient conditions is indeed
close to coexistence with its vapor. 

\subsection{Results}

\begin{figure}[t]
\epsfig{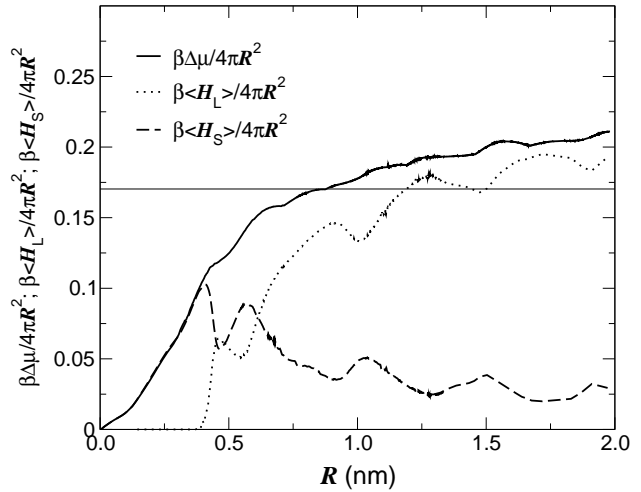}
\vspace*{0.4cm}
\caption{\label{fig:bdmu_E} The excess chemical potential per unit area of a hard sphere in
water as a function of its size; the hard sphere excludes water from a
spherical volume of radius $R$. The dotted line indicates the average
potential energy from the bare lattice gas model as given by
$H_{\rm L}[n]$ in Eq.~(\ref{eq:H_fs}) and the dashed line gives the
average potential energy of the solute-solvent interaction as given by
$H_{\rm S}[n]$ in Eq.~(\ref{eq:H_fs}). It is seen that the lattice
artifacts in the two energy contributions tend to cancel each
other. The horizontal line lies at the value of the surface tension
$\gamma$ of the vapor-liquid interface of water.}
\end{figure}

To test whether the theory successfully addresses density fluctuations at
all length scales, a good benchmark is the excess chemical potential of an
ideal solvophobic solute in a solvent as a function of its size~\cite
{Lum99,Huang00,Sun01}. Here we present results for the solvation of a hard
sphere (i.e., spherical bubble) in water.

The excess chemical potential $\Delta \mu $ of a hard sphere that excludes
 solvent from a region of volume $v_{\mbox{\tiny ex}}$, is
given by Eq.~(\ref{eq:bdmu}). We can obtain the excess chemical
 potential as a function of $v_{\mbox{\tiny ex}}$ by
sampling the size distribution of a ``breathing'' hard sphere with the
Hamiltonian $H[\left\{ n_{k}\right\}]$ shown in
Eq.~(\ref{eq:H_fs}). Specifically, in a Monte Carlo trajectory for the
large length scale field, each trial move consisted either of an attempt to
flip a spin or to change the radius of the solute; the trial moves are
accepted with a probability proportional to $\exp \left(-\beta \Delta H\right)$, where $\Delta H$ is the change in $H[\left\{ n_{k}\right\} ]$ due to the move. In
order to obtain accurate statistics for all solute sizes, we have used
umbrella sampling~\cite{Torrie74}.

Figure~\ref{fig:bdmu_E} shows the excess chemical potential of a hard
sphere as a function of $R$, where $R$ is the radius of the spherical
excluding volume. The results do exhibit some lattice artifacts. These
artifacts, however, are surprisingly small given the fact that the
cells are quite large. The broken lines in Fig.~\ref{fig:bdmu_E}
reveal why the lattice artifacts are small.  These curves show the
contribution to the solvation free energy from the energy of the
solvent, as given by $H_{{\rm L}}[\left\{ n_{k}\right\} ]$, and from
the energy of the ``solute-solvent'' interaction, as given by $H_{
{\rm S}}[v_{\mbox{\tiny ex}};\left\{ n_{k}\right\} ]$ in
Eq.~(\ref{eq:H_fs}).  Clearly, the discrete nature of our separation
of length scales is manifest in the strong energy changes at intervals
of length comparable to the grid spacing. This behavior is not
surprising, as the large length scale field, $n_{i},$ is effectively
very cold ($\epsilon =6.02k_{\rm B}T$). The important point to note is
that the lattice artifacts in the respective free-energy contributions
tend to properly cancel each other.

\begin{figure}[t]
\epsfig{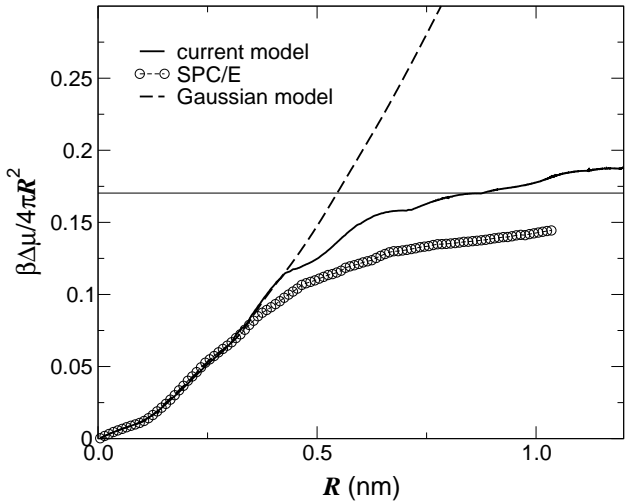}
\vspace*{0.45cm}
\caption{\label{fig:bdmu_SPCE}
Comparison of the results of the
present model with the predictions of the Gaussian model,
Eq.~(\ref{eq:bdmu_G}), and the results of a molecular simulation of a
cavity in SPC/E water [14].  The horizontal line lies at
the value of the surface tension $\gamma$ of the vapor-liquid
interface of water. }
\end{figure}

\begin{figure}[t]
\epsfig{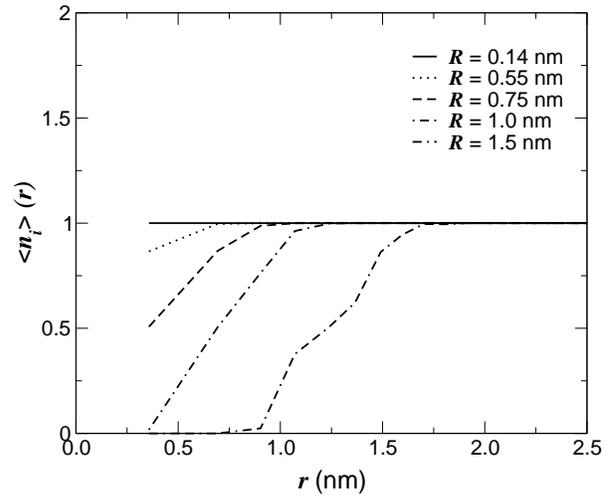}
\vspace*{0.1cm}
\caption{\label{fig:ni_r}  Slowly varying density $\langle n(r)\rangle $ for hard spheres of
different size, as a function of $r$, where $r$ is the distance to the
center of the solute. This radial profile was obtained by averaging
$n_i$ in concentric shells of radius $r$ and width $\Delta
r=0.1$\AA. It is seen that the small solutes are in the wetting
regime $\langle n(r)\rangle \approx \langle n(r)\rangle _{o}\approx 1.0$,
whereas the larger solutes are in the drying regime, for which $\langle
n(r)\rangle $ approaches a vapor-like value in the core of the solute.}
\end{figure}

In Fig.~\ref{fig:bdmu_SPCE} we compare the results of our model with the
predictions of the Gaussian model, Eq.~(\ref{eq:bdmu_G}), and with the
results of a molecular simulation of a hard sphere in SPC/E water~\cite
{Huang01_1}. It is seen that for small solute sizes, the agreement between
the results of the fluid models and the SPC/E-simulation results is very
good. The agreement is expected, since for small solutes, the large
length scale density
field remains close to its value in the unperturbed fluid with $\langle
n_{i}\rangle \approx 1$, as can be seen from the radial density profiles in
Fig.~\ref{fig:ni_r}. In this regime, our model reduces to the
Gaussian model, Eq.~(\ref{eq:H_G}). Computer simulations have shown
that at small length scales, density fluctuations in water obey Gaussian
statistics~\cite{Hummer96}. Thus, a Gaussian model and hence the
present model successfully predict the excess chemical potential of
small apolar species in water.

For solutes with $R>4$\AA , the predictions of the Gaussian theory
diverge from those of the full theory. The divergence is due to
drying, as revealed in Fig.~\ref{fig:ni_r}. The large length scale
field $\langle n_{i}\rangle $ approaches a vapor-like value in the
core of the larger solutes. Gaussian models cannot describe this
drying or depletion, because they are based upon a density expansion
around the uniform fluid.
In order to describe drying, a fluid model has to support
such a microscopic manifestation of a gas-liquid phase transition.

\begin{figure}[t]
\epsfig{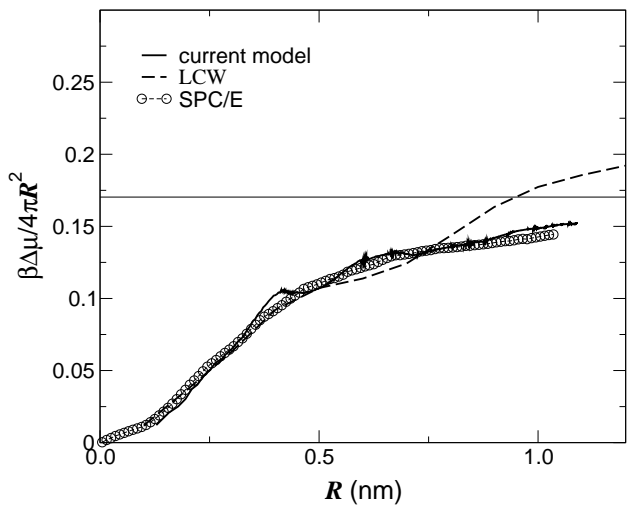}
\vspace*{0.25cm}
\caption{\label{fig:at1.5}  Excess chemical potential per unit area for a hard sphere in water at
ambient conditions (solid line). Here the energy density $a$ and the
interaction parameter $\epsilon^{\prime}$ are increased by fifty percent with
respect to the results shown in Fig.~\ref{fig:bdmu_SPCE}. The dashed line
denotes the results of the theory developed by Lum, Chandler, and
Weeks [1]. The molecular simulation results for a cavity in SPC/E water are
indicated by the circles [14].}
\end{figure}

One of the attractive features of the present model is that it lays
 bare the relative contributions to the solvation free energy. In the
 crossover regime, the contribution to the Hamiltonian from the
 unbalancing potential, is very important. When we increase $\epsilon
 ^{\prime }$ by fifty percent from that estimated by
 Eq.~(\ref{eq:a_epsp}), the results of the model agree very well with
 the simulation results over the entire range over which there is
 simulation data. In particular, compare Figs.~\ref{fig:bdmu_SPCE} and
 \ref{fig:at1.5}. Evidently, orientational
 degrees of freedom result in an unbalancing potential that is larger
 than that estimated for simple fluids. With the simple fluid estimate
 of the unbalancing potential, the LCW theory~\cite{Lum99}
 overestimates the excess chemical potential in the crossover
 regime. It appears that a somewhat larger unbalancing potential would
 correct this deficiency in LCW theory as it does in the current model.

It is often assumed that the excess chemical potential of an apolar species
in water is proportional to its exposed surface area. Our results, as well
as those of the LCW theory, emphasize that this is a reasonable assumption
in the drying regime. As the solvent is near phase coexistence, the
difference in chemical potential between vapor and liquid is small, and the
work done to insert a solute predominantly arises from the
work to create a vapor-liquid interface. For small solutes, however, the
excess chemical potential does not scale in this way. To a better
approximation, in this regime, it is a linear function of  $v_{\mbox{\tiny ex}}$, as can be seen from Figs.~\ref{fig:bdmu_SPCE} and
 ~\ref{fig:at1.5}.

\section{Discussion}

\label{sec:discussion} We have developed a new model for a cold liquid
that captures the effects of density fluctuations at both small and
 large length scales. This development is important, because many
 phenomena in liquids involve the interplay of density fluctuations in
 the two regimes. One example is capillary condensation. Another is
 hydrophobicity. Here we have focussed on the solvation of a
 hydrophobic species. Its nature
 is very different at small and large length scales. At small length
 scales, solvation is dominated by entropic
 effects. In this regime, the solvent can still wet the surface of the
 solute, even when the solute is highly
 hydrophobic~\cite{Lum99,Huang00}.  In contrast, at large length
 scales, solvation is dominated by energetic effects. In this regime,
 large hydrophobic objects can induce a drying transition in the
 solvent. Further, in the small length scale regime, the excess
 chemical potential scales with the volume of the solute, whereas in
 the large length scale regime, the excess chemical potential scales
 with the exposed area of the solute. The crossover behavior of the
 solvation free energy from the wetting to the drying regime would
 seem to be of significance to the self-assembly of biological
 structures~\cite{TenWolde01,Huang00_1}. In biological systems, the
 size of most hydrophobic species is such that individual species are
 in the wetting regime, while assemblies of such species are in the
 drying regime. Water can only induce a relatively weak attraction
 between two small hydrophobic species. When several of these species
 come together, however, water can induce a strong attraction
 between them.

The crossover behavior of the solvation free energy also implies that the
strength of the interactions between the hydrophobic species, depends on
the configuration of these species. The change in the interactions
manifests a collective effect in the solvent, and is therefore not
simply pair decomposable. Correct simulations of self assembly should
capture this collective effect, as could be done most
straightforwardly with an explicit solvent model. While atomistic
solvent models are highly limited for this purpose, the coarse-grained
model we have developed here should prove very useful~\cite{TenWolde01}.

The effects of weak interactions have been ignored in this
paper. Except for the movement of interfaces, that can be affected by
small forces, weak interactions
are not expected to induce large structural effects. Nevertheless,
their inclusion will be important for quantitative studies. The
inclusion of weak interactions can be accomplished by augmenting
Eq.~(\ref{eq:Xi}). Along with the constraint factor associated with the
hard core of the solute, the presence of a weak attractive potential,
$\phi({\bf r})$, between solute and solvent, will introduce the
additional factor $\exp\left[-\beta \int d{\bf r} \phi({\bf
r})\right]$. All the analysis carried out subsequent to
Eq.~(\ref{eq:Xi}) can be similarly performed in the presence of this
factor. Electrostatic interactions can also be incorporated, but with
somewhat greater complexity. In this case, liquid cells (with $n_i=1$), must
also possess a local polarization or dipole field, ${\bf m}_i$. This
vector field is Gaussian to a reasonable approximation~\cite{Song96}
and therefore can also be integrated out. These extensions of the
current model are left to future work.

\section*{Acknowledgments}
This work has been supported in its initial stages by the National
Science Foundation (Grant No. 9508336 and 0078458) and in its final
stages by the Director, Office of Science, Office of Basic Energy
Sciences, of the U.S. Department of Energy (Grant
No. DE-AC03-76SF00098).


\begin{thebibliography}{10}

\bibitem{Lum99}
{K. Lum, D. Chandler, and J. D. Weeks}, {\rm J. Phys. Chem. B} {\bf 103}, 4570
  {(1999)}.\vspace{-0.3pt}

\bibitem{Hummer96}
{G. Hummer, S. Garde, A. E. Garc\'{i}a, A. Pohorille, and L. R. Pratt}, {\rm
  Proc. Natl. Acad. Sci. USA} {\bf 93}, 8951 {(1996)}.\vspace{-0.3pt}

\bibitem{Crooks97}
G.~E. Crooks and D.~Chandler, {\rm Phys. Rev. E} {\bf 56}, 4217
  {(1997)}.\vspace{-0.3pt}

\bibitem{Percus58}
J.~K. Percus and G.~J. Yevick, {\rm Phys. Rep.} {\bf 110}, 1
  {(1958)}.\vspace{-0.3pt}

\bibitem{Wertheim63}
M.~S. Wertheim, {\rm Phys. Rev. Lett.} {\bf 10}, 321 {(1963)}.\vspace{-0.3pt}

\bibitem{Lebowitz66}
{J. L. Lebowtiz and J. K. Percus}, {\rm Phys. Rev.} {\bf 144}, 251
  {(1966)}.\vspace{-0.3pt}

\bibitem{Wertheim71}
M.~S. Wertheim, {\rm J. Chem. Phys.} {\bf 55}, 4291 {(1971)}.\vspace{-0.3pt}

\bibitem{Pratt77}
{L. R. Pratt and D. Chandler}, {\rm J. Chem. Phys.} {\bf 67}, 3683
  {(1977)}.\vspace{-0.3pt}

\bibitem{Chandler72}
{D. Chandler and H. C. Andersen}, {\rm J. Chem. Phys.} {\bf 57}, 1930
  {(1972)}.\vspace{-0.3pt}

\bibitem{Schweizer87}
{K. S. Schweizer and J. G. Curro}, {\rm Phys. Rev. Lett.} {\bf 58}, 246
  {(1987)}.\vspace{-0.3pt}

\bibitem{Chandler93}
{D. Chandler}, {\rm Phys. Rev. E} {\bf 48}, 2898 {(1993)}.\vspace{-0.3pt}

\bibitem{Weeks98}
{J. D. Weeks, K. Katsov, and K. Vollmayr}, {\rm Phys. Rev. Lett.} {\bf 81},
  4400 {(1998)}.\vspace{-0.3pt}

\bibitem{Chandler87}
D.~Chandler, {\em Introduction to Modern Statistical Mechanics}, Oxford
  University Press, New York {(1987)}.\vspace{-0.3pt}

\bibitem{Huang01_1}
{D. M. Huang, P. L. Geisller, and D. Chandler}, {J. Phys. Chem., in press
  (2001).}\vspace{-0.3pt}

\bibitem{Rowlinson82}
J.~S. Rowlinson and B.~Widom, {\em Molecular Theory of Capillarity}, Clarendon,
  Oxford {(1982)}.\vspace{-0.3pt}

\bibitem{note1TenWolde01}
{We have compared this approximation with seemingly more accurate
  approximations that include more basis sets. The comparisons indicate,
  however, that the approximation of one basis function per solute is
  accurate.}\vspace{-0.3pt}

\bibitem{TenWolde01}
{P. R. ten Wolde and D. Chandler}, {to be submitted}.\vspace{-0.3pt}

\bibitem{Hummer98}
{G. Hummer, S. Garde, A. E. Garc\'{i}a, M. E. Paulaitis, and L. R. Pratt}, {\rm
  J. Phys. Chem. B} {\bf 102}, 10469 {(1998)}.\vspace{-0.3pt}

\bibitem{Narten71}
{A. H. Narten and D.Levy}, {\rm J. Chem. Phys.} {\bf 55}, 2263
  {(1971)}.\vspace{-0.3pt}

\bibitem{note2TenWolde01}
{Relation~(\ref{eq:a_epsp}) can be understood as follows: The energy per unit
  volume in a uniform fluid of density $\rho_l$ is $a \rho_l^2$. The energy per
  unit volume in the lattice model is $z \epsilon^{\prime} / 2 l^3$. Equating
  the two yields Eq.~(\ref{eq:a_epsp}).}\vspace{-0.3pt}

\bibitem{Huang00}
D.~M. Huang and D.~Chandler, {\rm Phys. Rev. E} {\bf 61}, 1501
  {(2000)}.\vspace{-0.3pt}

\bibitem{Sun01}
{S. X. Sun}, {Phys. Rev. E, in press (2001).}\vspace{-0.3pt}

\bibitem{Torrie74}
G.~M. Torrie and J.~P. Valleau, {\rm Chem. Phys. Lett.} {\bf 28}, 578
  {(1974)}.\vspace{-0.3pt}

\bibitem{Huang00_1}
D.~M. Huang and D.~Chandler, {\rm Proc. Natl. Acad. Sci. USA} {\bf 97}, 8324
  {(2000)}.\vspace{-0.3pt}

\bibitem{Song96}
{X. Song, D. Chandler, and R. A. Marcus}, {\rm J. Phys. Chem} {\bf 100}, 11954
  {(1996)}.\vspace{-0.3pt}

\end{thebibliography}
\end{document}